\documentstyle[aps,prl]{revtex}  			
\begin{document}
\twocolumn[\hsize\textwidth\columnwidth\hsize\csname  	 
@twocolumnfalse\endcsname				
\draft
\title{Realistic Surface Scattering and Surface Bound State Formation in the 
High T$_c$ Superconductor YBa$_2$Cu$_3$O$_{6+x}$}
\author{M. B. Walker and P. Pairor }
\address{Department of Physics, 
University of Toronto,
Toronto, Ont. M5S 1A7 }
\date{\today }
\maketitle

\widetext					%
\begin{abstract}
Surface Umklapp scattering of quasiparticles, and surface 
roughness are shown to play essential roles in the formation of the surface 
bound states in realistic models for YBa$_2$Cu$_3$O$_{6+x}$.  
The results account for the shape, the impurity dependence of the height, and for a proposed universal width of the zero bias conductance peak.
\end{abstract}


\vfill		
\narrowtext			%

\vskip2pc]	


This article describes the formation of 
mid-gap surface bound states in models of YBa$_2$Cu$_3$O$_{6+x}$ (YBCO) 
which include realistic surface scattering of the quasiparticles.  
(We define realistic surface scattering to be scattering that includes surface 
roughness as well as surface Umklapp scattering processes and a realistic 
Fermi surface geometry.)  More than a trivial extension of what is known 
about surface bound states from the study of elementary models 
\protect\cite{hu94,tan95,mat95,buc95,xu96,yam96,fog97,zhu98,wal98} is 
required. (By our definition, elementary models neglect either surface 
roughness or surface Umklapp scattering or both).  On the one hand, except 
at grazing angles, the superconducting quasiparticles are prevented from 
forming surface bound states because they are reflected diffusely at a rough 
surface, and on the other hand, the grazing incidence quasiparticles are 
prevented from forming surface bound states on a [110] surface (the most 
favorable case) when surface Umklapp scattering is considered 
\protect\cite{wal98}.  Surface bound states are of interest in the study of 
high T$_c$ superconductors, because they probe the sign changes of the gap 
function $\Delta({\bf k})$ on the Fermi surface, and also because they 
manifest themselves experimentally through the zero bias conductance peak 
(ZBCP) observed in tunneling experiments.  As noted in the conclusions, our 
results have a considerable impact on the interpretation of experiments.

The existence \protect\cite{hu94} of surface bound states in $d$-wave 
superconductors and their role in producing the ZBCP observed in tunneling 
experiments 
\protect\cite{gee88,les92,kas94,cov96,alf97,eki97,apr98,wei98,wei98a} are 
well established for elementary models, e.g. in \protect\cite{tan95}.  Further 
studies have discussed, among other results, the existence of multiple sub-gap resonances \protect\cite{xu96}, the surface-induction of order 
parameters of different symmetries \protect\cite{mat95,buc95,fog97,sig95} 
and the associated splitting of the ZBCP at low temperatures 
\protect\cite{fog97,zhu98,wal98,cov96}, the splitting of the 
ZBCP in a magnetic field \protect\cite{fog97,cov96}, the effects of irradiation 
in suppressing the ZBCP \protect\cite{apr98}, the effects of surface 
roughness in broadening the ZBCP \protect\cite{mat95,yam96,fog97,tan98}, 
and the anisotropy of the ZBCP \protect\cite{wei98a}.

According to the Rayleigh criterion in the theory of surface roughness (for a 
general discussion see \protect\cite{vor94}), a surface is rough 
(reflects diffusely) or smooth (reflects specularly) according as the 
parameter $R_k = k_\perp \eta$ is smaller than or greater than unity.  Here 
$k_\perp$ is the component of the incident-wave wave vector perpendicular 
to the surface and $\eta$ is the surface asperity.  For $R_k$ 
small, perturbation theory (e.g. see 
\protect\cite{vor94}) gives the 
probability of diffuse reflection as $P_D  \sim R_k^2$, i.e. 
it goes to zero as $k_\perp^2$ for small $k_\perp$.  On the other hand, for 
$R_k$ large, a quasiclassical approximation shows that, for a wave incident 
normally on the surface, the probability of 
specular reflection is given approximately by $P_S = exp(-4 \pi R_k^2)$, 
i.e. there is essentially no specular reflection for $R_k$ greater than unity 
(e.g. see \protect\cite{zim60}).  

It follows from the above estimates that, except for quasiparticles approaching 
the surface at grazing incidence, the reflection is totally diffuse.  These ideas 
have been successfully applied to the theory of the electronic surface states 
of a normal metal in low magnetic fields \protect\cite{pra68} where  
microwave absorption experiments show that quantized magnetic surface 
levels exist only for electrons incident on the surface at grazing angles (so 
that they are specularly reflected).  The application of these ideas to
surface bound states in d-wave superconductors suggests that, except 
possibly for those which approach the surface at grazing angles, the 
quasiparticles will be diffusely reflected at a rough surface and thus will not 
maintain the phase coherence necessary (e.g. see the argument surrounding 
Eq. 25 of \protect\cite{tan95}) to form bound states.  

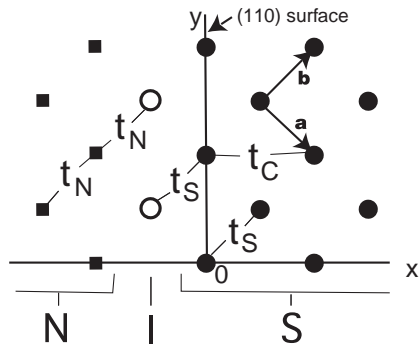
\begin{figure}
\caption{A model (110) NIS tunnel junction.  The solid (open) circles, and the 
solid squares represent ions in the superconductor (insulator), and the 
normal metal, respectively.  Also, $t_N$  is the nn normal metal (and normal 
to insulating) hopping interaction, while $t_S$ and $t_C$ are the nn and nnn 
superconductor hopping interactions.  We also assume a nn d-wave gap 
function, i.e. $\Delta_a = - \Delta_b$.}
\label{fig1}
\end{figure}

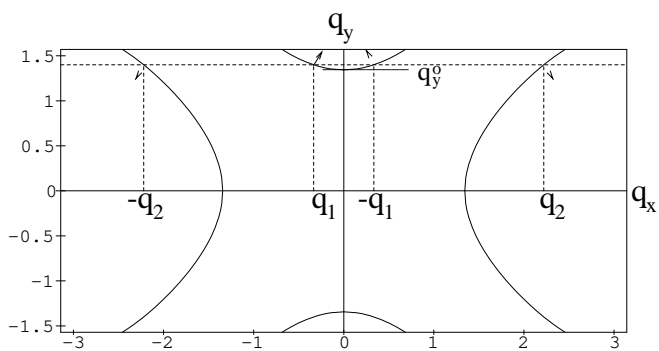
\begin{figure}
\caption{The surface-adapted Brillouin zone and Fermi surface for a (110) 
surface.  The notation $q_{x,y}=k_{x,y}a/\sqrt{2}$ is used.  The Fermi surface 
shown corresponds qualitatively to that measured experimentally (e.g. see 
\protect\cite{wal98}).  Note that $q_1<0$.}
\label{fig2}
\end{figure}

Surface bound states in a model which incorporates surface Umklapp 
scattering have recently been 
described in \protect\cite{wal98}.  As in \protect\cite{wal98,odo95}, to 
obtain a realistic Fermi surface \protect\cite{din96,sch97} we use a discrete 
lattice model (see Fig.\ \ref{fig1}) including 
both nearest and next nearest neighbor hopping interactions in the 
superconductor.  Also, we consider the case of a [110] surface, for which the 
formation of surface bound states is the most 
favorable.  The surface bound states are eigenstates of the two-component 
Bogoliubov-de Gennes equations and are linear combinations of 
basic solutions of the form
\begin{equation}
	U_{\bf k_i}({\bf x}) = 
			\left[ 
			\begin{array}{c}
				\Delta_i \\
				E \pm i\Omega_i	
			\end{array} \right]
e^{i(k_{ix}x + k_y y)}
		e^{-\kappa_i x}.
	\label{eq1}
\end{equation}
as in \protect\cite{hu94,tan95,xu96,wal98} and others.
Here ${\bf k_i}= (k_{ix},k_y)$ must be on the Fermi surface, 
$\Delta_i = \Delta({\bf k_i})$, $\Omega_i = \sqrt{\Delta_i^2 - E^2}$, 
and $\kappa_i  = \Omega_i /(\hbar |v_{ix}|)$ with $v_{ix}$
being the $x$ component of the normal-state electron velocity.  The upper 
and lower signs in Eq.\ \ref{eq1} correspond to 
$ v_{ix} < 0 $ and $v_{ix} > 0$,
respectively.  From Fig.\ \ref{fig2} it is seen that for $k_y > k_y^0$, a line of 
constant $k_y$ intersects the Fermi surface at four 
distinct points, giving four values $k_{ix}$, i= 1 ... 4, and hence 
four linearly independent exponentially decaying solutions. In this case the 
surface bound state solutions have the form
\begin{equation}
	U({\bf x}) = \Sigma_i C_i U_{\bf k_i}({\bf x}).		
	\label{eq2}
\end{equation}

From above, only grazing incidence quasiparticles, i.e. those 
corresponding to small values of $|k_{1x}|$ in the surface-adapted Brillouin 
zone \protect\cite{wal98} of Fig.\ \ref{fig2}, will have the possibility to form 
surface bound states in the case of rough surfaces.  
However, it turns out that a detailed study of such states 
\protect\cite{wal98} has shown that, at least in the case of perfectly flat 
surfaces, waves having $k_y > k_y^0$ in Fig.\ \ref{fig2}  (and this 
includes all waves of small $|k_{1x}|$) can  not form surface bound states  
[the reason has to do with the relative signs of the gaps $\Delta(k_{1x}, k_y)$ 
and $\Delta(k_{2x}, k_y)$].

In summary, on the one hand, except at grazing incidence, quasiparticles (or 
holes) striking the surface are reflected diffusely and hence do not have 
the coherence necessary to form surface bound states, while on the 
other hand, grazing incidence quasiparticles are prevented from forming bound 
states in a realistic model for the Fermi surface because they do not satisfy 
the necessary conditions on the sign variation of the gap $\Delta({\bf k})$.   

We now give a qualitative explanation of the result of our quantitative 
calculation presented below showing how it is possible 
to form surface bound states in models of YBCO which include realistic 
surface scattering.  First, assume that the surface at $x=0$ is perfectly flat,  
and consider an incoming wave of the form of Eq.\ \ref{eq1} with (the x-
component of its) wave vector equal to $-k_{1x}$, or equivalently $-q_1$, of 
Fig.\ \ref{fig2}.  In order to satisfy the boundary conditions at the surface, 
two outgoing waves, a specularly scattered wave at $k_{1x}$ and an 
Umklapp scattered wave at $k_{2x}$, are required.  (Note that the wave at 
$k_{2x}$ appears as a normal scattering process when viewed in the surface-
adapted Brillouin zone of Fig.\ \ref{fig2}, but is a surface Umklapp process 
when viewed in the normal bulk Brillouin zone \protect\cite{wal98}.)
It is the Umklapp scattering into the state with wave vector $k_{2x}$ which 
prevents the formation of the surface bound state because the relative signs 
of the gaps at wave vectors $\pm k_{1x}$ and $\pm k_{2x}$ are 
incompatible with the formation of a surface bound state 
\protect\cite{wal98}.  Thus we can only get surface bound states (or at least 
resonances) if the scattering from  wave vectors $\pm k_{1x}$ to $\pm 
k_{2x}$ is sufficiently reduced.

It is the roughness of the surface which limits the Umklapp scattering of 
quasiparticles of 
wave vectors $\pm k_{1x}$ into the states of wave vector $\pm k_{2x}$. 
Quasiparticles in states which are  linear combinations of states with wave 
vectors $\pm k_{2x}$ will be strongly diffusely reflected by a rough surface 
and will thus have short lifetimes.  Their short lifetimes will spread out their 
density of states in energy, reducing the density of states at the energy of 
the quasiparticles of wave vector $k_{1x}$ and hence reducing the probability of 
scattering from wave vectors $\pm k_{1x}$ to $\pm k_{2x}$.

We now develop a more quantitative theory of the effects of surface 
roughness by assuming that surface-bound-state component states (Eq.\ 
\ref{eq1}) having wave vectors $\pm k_{ix}$ have a finite lifetime $\tau_i = 
\hbar/\gamma_i$ due to nonspecular scattering at the surface.  For quasiparticles 
with vectors $\pm k_{2x}$  (which have a probability of unity for diffuse 
scattering at the surface) we take this lifetime to be the time taken for an 
electron to get to the surface from a distance of $(\kappa_i)^{-1}$, which, for 
$|E| \ll |\Delta_2|$ gives $\gamma_2 = |\Delta_2|$.  (Similar qualitative 
estimates of damping rates were shown in \protect\cite{wal98} to be in 
excellent agreement with the results obtained from a detailed solution of the 
Bogoliubov-de Gennes equations.) For grazing incidence quasiparticles of wave 
vector $\pm k_{1x}$, which have a very small probability $P_1$ of diffuse 
scattering at the surface, the appropriate value of $\gamma$ is $\gamma_1 
= P_1 |\Delta_1|$. (The perturbation theory estimate of $P_1$ is given in the 
above general discussion of surface roughness.)  These lifetime effects are 
incorporated into our model by using the phenomenological approach 
(\protect\cite{fog97,alf97}) of replacing $E$ by $E+i\gamma_i$ in Eq.\ 
\ref{eq1} for $U_{\bf k_i}$.  If we now look for solutions of the form of Eq.\ 
\ref{eq2} which satisfy the appropriate boundary conditions for a 
superconductor to vacuum surface (corresponding to the absence of the 
normal metal and the insulating layer in Fig.\ \ref{fig1}), and which are 
valid in the limit of small $|k_{1x}|$, we find a solution having an energy $E = 
E_B + i\Gamma_R$  where the bound state energy is $E_B = 0$ and the width 
due to roughness is $\Gamma_R =\gamma_1 + \Gamma_{12}$, with the 
component 
\begin{equation}
	\Gamma_{12}(k_y) = |\Delta_1|(\Upsilon_2 + |\Delta_2|)|sin q_1 
	cot(q_2/2)|/\gamma_2.
	\label{eq3}
\end{equation}
Here $\Upsilon_2 = \sqrt{\Delta_2^2 + \gamma_2^2}$ and $q_i = 
k_{ix}a/\sqrt{2}$.  Note that $\Gamma_{12}$ depends on $k_y$ through 
the dependences of $q_i$ and $\Delta_i$ on $k_y$.  From the above result 
$\Gamma_R = \gamma_1 + 
\Gamma_{12}$, we interpret the solution as being 
primarily a state formed from the grazing incidence wave vectors $\pm 
k_{1x}$ and having a width $\gamma_1$  which is augmented by 
$\Gamma_{12}$ due to transitions from wave vectors $\pm k_{1x}$ to $\pm 
k_{2x}$.  It is clear that for $\gamma_2$  having the value $|\Delta_2|$ 
suggested above, the width $\Gamma_{12}$ is smaller than $|\Delta_1|$ 
(since $q_1$ is small) and the bound state thus has a spread of energies 
lying within the gap  $|\Delta_1|$.  If $\gamma_2$ were to be small on the 
other hand, $\Gamma_{12}$ would be large and no well defined state would 
exist in the gap.  Clearly it is the broadening of the quasiparticles at wave 
vectors $\pm k_{2x}$  by diffuse scattering that allows the formation of 
surface bound states for the case of realistic surface scattering.

We now proceed to a calculation of the tunneling conductance for the model 
NIS junction illustrated in Fig.\ \ref{fig1} using a procedure well-established 
in other work, i.e. we use the BTK formalism \protect\cite{blo82} but 
extended to the case of d-wave superconductors 
\protect\cite{tan95,xu96,zhu98} and to a discrete-lattice model 
\protect\cite{wal98,tan98}.  The procedure is straightforward: assume a 
solution in the form of a linear combination of an electron moving towards 
the junction together with a reflected electron and reflected hole in normal 
state, and a linear combination of the form of Eq.\ \ref{eq2} in the 
superconducting state.  The coefficients in the linear combinations are found 
using the boundary condition implicit in Fig.\ \ref{fig1} and the tunneling 
conductance is calculated following BTK \protect\cite{tan95,blo82}. We 
restrict ourselves to the low temperature limit, and to the calculation of the 
contribution of the surface 
bound states to tunneling conductance, which is valid only for voltages such 
that eV is less than the maximum gap, and we also work in the weak 
transmission limit where the probability of an electron tunneling across the 
barrier is much less than unity, since it is only in this limit that one obtains a 
sharply defined ZBCP in the gap \protect\cite{xu96,wal98}.  In this way, we 
find that the conductance per surface unit cell (for eV smaller than the 
maximum gap) is given by the formula
\begin{equation}
	G=\frac{4e^2}{h} \left\langle \Gamma_{SN}(k_y)
		\frac{\Gamma_T(k_y) }{(eV)^2 + (\Gamma_T(k_y))^2} 				\right\rangle
	\label{eq4}
\end{equation}
where the angular brackets indicate an average over $k_y$.  Here
\begin{equation}
	\Gamma_T(k_y) = \gamma_1(k_y) + \gamma_{imp} + 
	\Gamma_{12}(k_y) + 	\Gamma_{SN}(k_y) 
	\label{eq5}
\end{equation}
where the damping rate of the basics states with wave vectors $\pm 
k_{1x}$, 
called $\gamma_1$ above, has been replaced by $\gamma_1 + 
\gamma_{imp}$, thus separating the momentum-dependent diffuse surface 
scattering part $\gamma_1$ (proportional to $q_1^3$) from the momentum-
independent 
bulk impurity scattering part, $\gamma_{imp}$.  Also
\begin{equation}
	\Gamma_{SN} = 2\alpha_N \alpha_C sin q (sin q_2 \gamma_1 
	\Upsilon_2 - sin q_1 |\Delta_1| \gamma_2)/\gamma_2,
	\label{eq6}
\end{equation}
with $\alpha_N = t_N/V_0$, $\alpha_C = t_C/V_0$, $V_0$ being the 
additional potential on the ions in the insulating layer I of Fig.\ \ref{fig1}, 
and $q>0$ being the wave vector of the incident normal state electron.  Our 
calculations are valid to the lowest nontrivial order in $\alpha_N$, 
$\alpha_C$ (i.e. a weak transmission insulating barrier) which means that 
the contribution of $\Gamma_{SN}$ to $\Gamma_T$ should be neglected (but 
that $\Gamma_{SN}$ should remain in the numerator).  The fact that 
$\Gamma_{SN}$ is nevertheless one of the contributions to $\Gamma_T$ 
allows its interpretation as a contribution to the lifetime of the surface 
bound 
state resulting from the tunneling of the bound state excitation from the 
superconductor into the normal metal.

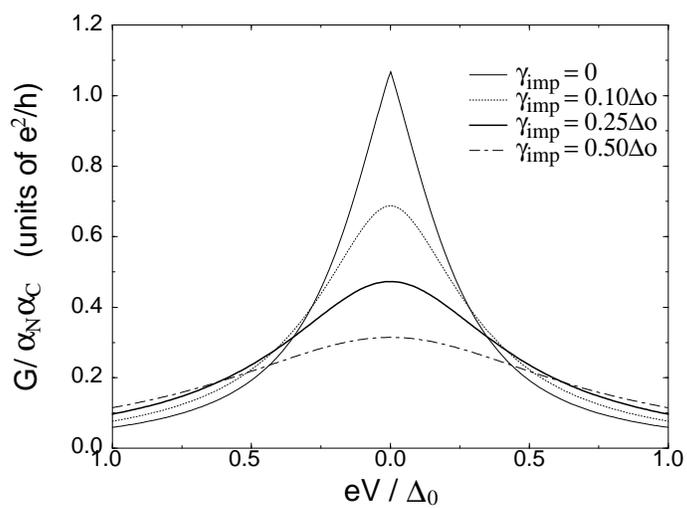
\begin{figure}
\caption{The surface bound state contribution to ZBCP for a (110) NIS 
junction as predicted by Eq.\ \ref{eq4}.}
\label{fig3}
\end{figure}

The tunneling conductance described by Eq.\ \ref{eq4} is plotted in Fig.\ 
\ref{fig3} as a function of $eV/\Delta_0$ (charge times voltage over 
maximum gap).  The contribution of $\gamma_1$ to $\Gamma_T$ is 
neglected since it is smaller than $\Gamma_{12}$ at small wave vectors 
$q_1$.  Thus the height, width and shape of the curve of conductance versus 
voltage are determined by $\Gamma_{12}$, $\Gamma_{SN}$, and 
$\gamma_{imp}$.  In particular, the cusp-shaped maximum at $eV$=0 in the 
curve for $\gamma_{imp} =0$ is due to the fact that $\Gamma_{12}$ and 
$\Gamma_{SN}$ both go to zero as $q_1^2$ at small $q_1$.  This means that 
the bound states with very small $q_1$ are very narrow and contribute all 
of their weight near $eV=0$, giving a sharp maximum there.  The broader 
bound states at larger values of $q_1$ give broader contributions to the 
conductance and are responsible for the weight in the wings.  A cusp shaped 
ZBCP has been observed in \protect\cite{wei98a}. 

A strong reduction of the ZBCP height with increasing disorder (i.e. increasing 
$\gamma_{imp}$) without, however, a significant increase in the ZBCP width 
has been found in \protect\cite{apr98}.  This agrees with the behavior which 
is predicted by our Eq.\ \ref{eq4} and shown in Fig.\ \ref{fig3}.  The 
qualitative explanation for this can be found in the momentum dependence 
of $\Gamma_{12}$.  The very small $q_1$ quasiparticles, which are very 
narrow in energy, give a high ZBCP in the absence of impurity scattering.  
Impurity scattering broadens these quasiparticles and reduces their 
contribution to the peak height, but has little effect on the broader 
quasiparticles at less small $q_1$ which are responsible for the overall width of 
the ZBCP.

Eq. (4) predicts a universal width, independent of surface asperity and 
impurity scattering (for impurity scattering not too strong), for the ZBCP 
assuming that $\Gamma_T$ is dominated by $\Gamma_{12}$.  From Fig.\ 
\ref{fig3}, this universal half width at half maximum is approximately 
$\Delta_0/4$.  This is very different from the predictions of elementary 
models where the width is due to and increases in proportion to the surface 
asperity \protect\cite{mat95,yam96} or impurity scattering 
\protect\cite{fog97,alf97}.  Our result is roughly in agreement with 
experiment where the widths of measurements made under widely differing 
conditions (for an asperity $\eta \sim$ 1 nm in \protect\cite{alf97}, for 
$\eta \sim$ 1 to 10 nm in \protect\cite{tan95}  for $\eta \sim$ 500 nm. in 
\protect\cite{wei98a}, and for varying $\gamma_{imp}$ in 
\protect\cite{apr98}) all give aproximately the same half width of about 2 to 2.5 meV, which is 
however about a factor of two smaller than our prediction.  The error in the 
numerical factor is perhaps due to our simplistic representation of the 
insulating barrier by a single atomic layer, or to our simplistic treatment of 
the surface roughness (carried out simply by the replacement of $E$ by 
$E+i\gamma_i$).

Our results, in which the width of the ZBCP is due to $\Gamma_{12}$, also 
resolve the problem that fitting the ZBCP to elementary models requires an 
inexplicably large value of the so-called smearing factor [see 
\protect\cite{tan95}(a)]. 

To conclude, we note that the description of surface bound 
state formation and the consequent effects on the tunneling conductance in 
YBCO requires a realistic description of the surface scattering processes. The 
principal source of width of the surface bound states is their limited lifetime 
due to the surface Umklapp scattering from grazing incidence states to larger 
perpendicular momentum states, and this process hence has a determining 
effect on the height, width and shape of the ZBCP.  These ideas give a 
qualitative explanation of the cusp shaped ZBCP observed in 
\protect\cite{wei98a}, the height reduction (without width increase) of the 
ZBCP caused by increased disorder seen in \protect\cite{apr98}, and the 
approximately universal value (independent of surface roughness and 
impurity concentration) of the width of the ZBCP observed in 
\protect\cite{tan95,cov96,alf97,apr98,wei98a}. 

A useful discussion with M. Aprili is acknowledged, as are the hospitality of 
P. Nozi\`{e}res and the Institut Laue Langevin Theory Group (where part of 
this work was done) and the support of the Natural Sciences and Engineering 
Research Council of Canada.

\begin{references}

\bibitem{hu94} C. R. Hu, Phys. Rev. Lett. {\bf 72}, 1526 (1994); Phys. Rev. B 
{\bf 57}, 1266 (1998).

\bibitem{tan95} Y. Tanaka and S. Kashiwaya, Phys. Rev. Lett. {\bf 74}, 3451 
(1995); S. Kashiwaya {\it et al.}, Phys. Rev. B {\bf 51}, 1350 (1995); {\bf 53}, 
2667 (1996); (a) {\bf 57}, 8680 (1998).

\bibitem{mat95} M. Matsumoto and H. Shiba, J. Phys. Soc. Jpn. {\bf 64}, 1903 
(1995); {\bf 64}, 3384 (1995); {\bf 64}, 4867 (1995).

\bibitem{buc95} L. Buchholtz {\it et al.}, J. Low Temp. Phys. {\bf 101}, 1099 
(1995); {\bf 101}, 1079 (1995).

\bibitem{xu96} J. H. Xu {\it et al.}, Phys. Rev. B {\bf 53}, 3604 (1996).

\bibitem{yam96} K. Yamada et al.,  J. Phys. Soc. Jpn. {\bf 65}, 1540 (1996). 

\bibitem{fog97} M. Fogelstr\"{o}m {\it et al.}, Phys. Rev. Lett. {\bf 79}, 281 
(1997); Rainer {\it et al.}, cond-mat/9712234.

\bibitem{zhu98} J. Zhu and C. S. Ting, Phys. Rev. B {\bf 57}, 3038 (1998).


\bibitem{wal98} M. B. Walker and P. Pairor, cond-mat/9803079.

\bibitem{gee88} J. Geerk {\it et al.} Z. Phys. B {\bf 73}, 329 (1988).

\bibitem{les92} J. Lesueur {\it et al.}, Physica C {\bf 191}, 325 (1992).

\bibitem{kas94} S. Kashiwaya {\it et al.}, Physica B {\bf 194-196}, 2119 
(1994).

\bibitem{cov96} M. Covington {\it et al.}, Appl. Phys. Lett. {\bf 68}, 1717 
(1996); Phys. Rev. Lett. {\bf 79} 277 (1997).

\bibitem{alf97} L. Alff {\it et al.}, Phys. Rev. B {\bf 55}, 14757 (1997).

\bibitem{eki97} J. W. Ekin {\it et al.}, Phys. Rev B {\bf 56}, 13746 (1997).

\bibitem{apr98} M. Aprili, M Covington, E. Paraoanu, B. Niedermeier, 
and L. H. Greene, Phys. Rev. B {\bf 57}, R8139 (1998).

\bibitem{wei98} J. Y. T. Wei, Phys. Rev. B {\bf 57}, 3650 (1998).

\bibitem{wei98a} J. Y. T. Wei, N.-C. Yeh, D. F. Garrigus, and M. Strasik, 
preprint.

\bibitem{sig95} K. Kuboki and M. Sigrist, J. Phys. Soc. Jpn. {\bf 65}, 361 
(1996); M. Sigrist {\it et al.}, Phys. Rev. Lett. {\bf 74}, 3249 (1995).

\bibitem{tan98} Y. Tanuma, Y. Tanaka, M. Yamashiro, and S. Kashiwaya, 
Phys. Rev. B {\bf 57}, 7997 (1998).

\bibitem{vor94} A. G. Voronovich, {\it Wave Scattering from Rough 
Surfaces,} \ Springer-Verlag (Berlin Heidelberg 1994).

\bibitem{zim60} J. M. Ziman, {\it Electrons and Phonons}, Oxford University 
Press (1960).

\bibitem{pra68} R. E. Prange and T-W. Nee, Phys. Rev. {\bf 168}, 779 (1968).

\bibitem{odo95} C. O'Donovan and J. P. Carbotte, Phys. Rev. B {\bf 52}, 4548 
(1995).

\bibitem{din96} H. Ding {\it et al.}, Phys. Rev. B {\bf 54}, R9678, (1996).

\bibitem{sch97} M. C. Schabel {\it et al.}, Phys. Rev. B {\bf 55}, 2796 (1997).

\bibitem{blo82} J. Demers and A. Griffin, Can. J. Phys. {\bf 49},
285 (1970); G. E. Blonder, M. Tinkham, and T.M Klapwijk,
Phys. Rev. B {\bf 25}, 4515 (1982).

\end{references}
\end{document}